# Grid-free powder averages: on the applications of the Fokker-Planck equation to solid state NMR

Luke J. Edwards[a,b], D.V. Savostyanov[b], A.A. Nevzorov[c],
M. Concistre[b], G. Pileio[b], Ilya Kuprov[b,*]

[a]*Inorganic Chemistry Laboratory, Department of Chemistry,
University of Oxford, South Parks Road, Oxford, OX1 3QG, UK*

[b]*School of Chemistry, University of Southampton,
Highfield Campus, Southampton, SO17 1BJ, UK*

[c]*Department of Chemistry, North Carolina State University,
2620 Yarbrough Drive, Raleigh, NC 27695, USA*

Fax: +44 2380 594140

Email: i.kuprov@soton.ac.uk

**Abstract**

We demonstrate that Fokker-Planck equations in which spatial coordinates are treated on the same conceptual level as spin coordinates yield a convenient formalism for treating magic angle spinning NMR experiments. In particular, time dependence disappears from the background Hamiltonian (sample spinning is treated as an interaction), spherical quadrature grids are avoided completely (coordinate distributions are a part of the formalism) and relaxation theory with any linear diffusion operator is easily adopted from the Stochastic Liouville Equation theory. The proposed formalism contains Floquet theory as a special case. The elimination of the spherical averaging grid comes at the cost of increased matrix dimensions, but we show that this can be mitigated by the use of state space restriction and tensor train techniques. It is also demonstrated that low correlation order basis sets apparently give accurate answers in powder-averaged MAS simulations, meaning that polynomially scaling simulation algorithms do exist for a large class of solid state NMR experiments.

# 1. Introduction

Fokker-Planck equations that describe the time evolution of probability distributions[1,2] were first used in a magnetic resonance context in the 1970s as part of the formalism that became known as the Stochastic Liouville Equation (SLE)[3-6]. To this day SLE remains the most general relaxation theory in magnetic resonance – it is non-perturbative and works at all magnetic fields and correlation times, from extreme narrowing to the solid limit[6-8].

Initial adoption of Fokker-Planck equations was complicated by large matrix dimensions[4-6] and subsequent work had to rely heavily on sparse matrix libraries and Lanczos techniques[3], but the exponential rise in computing power in the following 40 years has removed the problem – the same calculations take just a few seconds on a contemporary workstation. Modern programming languages have also made things easier – a *Matlab* implementation of SLE in our *Spinach* library[9] takes about a hundred lines.

The advances in computing power make the Fokker-Planck formalism worth re-visiting. One particular feature that to our knowledge remains unexplored in magnetic resonance is the possibility of treating *any* spatial dynamics that can be generated by a linear operator – the formalism itself is in no way restricted to diffusion. We demonstrate below that magic angle spinning simulations in particular stand to benefit – not only is the time dependence removed from the MAS Hamiltonian (something that may already be achieved using Floquet theory[10-13]), but the need for a spherical quadrature grid also disappears: the Fokker-Planck formalism solves directly for orientation distributions and obtains the powder average in a single run. Given the amount of work that has gone into two- and three-angle spherical grids[14-21] and the computational expense of three-angle averaging, the prospect of having a *grid-free* formalism is quite attractive. Importantly, the Fokker-Planck MAS formalism contains Floquet MAS theory as a special case – when the lab space motion operator is set to be rotation and the lab space basis is chosen to be complex exponentials of the rotation angle, the equations of Floquet theory are obtained.

Another significant open question in solid state NMR is about the possibility of using state space restriction techniques[9,22,23] that only include low orders of spin correlation and result in very significant acceleration of many types of spin dynamics simulations[9,24,25]. The Fokker-Planck formalism described below generalizes this question and makes it possible to ask whether *some correlations between spin and spatial degrees of freedom might also be* *redundant*. The answer to this question is shown below to be affirmative – it is demonstrated that low correlation order basis sets do provide accurate answers in powder-averaged MAS NMR simulations, although it is not entirely clear at the moment why they work so well.

## 2. Fokker-Planck equation for a spinning NMR sample

For a given spin system, the Liouville - von Neumann equation describing the dynamics is a linear partial differential equation for the density operator $\hat{\rho}(t)$:

$$\frac{\partial}{\partial t}\hat{\rho}(t) = -i\hat{\hat{H}}(\vec{x},t)\hat{\rho}(t) \tag{1}$$

in which $\hat{\hat{H}}(\vec{x},t)$ is the spin Hamiltonian commutation superoperator that can depend on the system orientation and/or conformation $\vec{x}$ in the lab space, which is also in general time-dependent. In deterministic cases Equation (1) is solved directly using matrix exponentials[26,27], and in stochastic cases a relaxation superoperator is first obtained using one of the available relaxation theories[27,28]. In the cases where an average over spin system orientations or conformations is required, it is obtained by numerical or semi-analytical integration over a discrete grid of points[14-21]. This is currently the standard setting and *modus operandi* in magnetic resonance simulations.

The Fokker-Planck formalism looks at the system from a different perspective – instead of the dynamics of the density matrix $\hat{\rho}(t)$ of a specific system travelling along $\vec{x}(t)$ it considers the dynamics of the joint probability density $p(\vec{x},\hat{\rho},t)$ of systems at different lab space points $\vec{x}$ in different spin states $\hat{\rho}$. In this picture both $\vec{x}$ and $\hat{\rho}$ are time-independent coordinates and the probability density flows through their space. In the absence of stochastic processes, the Fokker-Planck (*aka* Smoluchowski) equation of motion for $p(\vec{x},\hat{\rho},t)$ is:

$$\frac{\partial p(\vec{x},\hat{\rho},t)}{\partial t} = -\text{div}_{\hat{\rho}}\left[p(\vec{x},\hat{\rho},t)\hat{v}_{\hat{\rho}}\right] - \text{div}_{\vec{x}}\left[p(\vec{x},\hat{\rho},t)\vec{v}_{\vec{x}}\right] \tag{2}$$

It has a simple physical meaning – the change in probability density at a given point is equal to the divergence of its flux. The equation is quite generic – its exact form depends on the expressions for the velocities $\hat{v}_{\hat{\rho}}$ and $\vec{v}_{\vec{x}}$ in the spin space (defined as the space of all spin operators) and the lab space (which could be bigger than $\mathbb{R}^3$ if multiple rotational and conformational degrees of freedom are considered) respectively. For a rotating NMR sample, the velocity at a point $\hat{\rho}$ in spin space is given by Equation (1):

$$\hat{v}_{\hat{\rho}} = -i\hat{\hat{H}}(\vec{x},t)\hat{\rho} \tag{3}$$

It depends parametrically on the lab space coordinates because the spin Hamiltonian may be orientation-dependent. The lab space velocity at a specific point $\vec{x}$ may be obtained from the equations of classical mechanics describing rotational motion in three dimensions:

$$\vec{x}(t) = \exp\left[-i\hat{S}t\right]\vec{x}(0) \qquad \hat{S} = \omega_S\left(n_X \hat{L}_X + n_Y \hat{L}_Y + n_Z \hat{L}_Z\right)$$
$$\frac{d\vec{x}(t)}{dt} = -i\hat{S}\exp\left[-i\hat{S}t\right]\vec{x}(0) = -i\hat{S}\vec{x}(t) \quad \Rightarrow \quad \vec{v}_{\vec{x}} = -i\hat{S}\vec{x} \tag{4}$$

where $\{\hat{L}_X, \hat{L}_Y, \hat{L}_Z\}$ are infinitesimal generators of the rotation group and $\omega_S$ is the spinning frequency around the axis specified by a normalized $[n_X, n_Y, n_Z]$ vector. In the case of magic angle spinning $\hat{S} = (\hat{L}_X + \hat{L}_Y + \hat{L}_Z)/\sqrt{3}$. Note that the spatial motion operator $\hat{S}$ is assumed to be classical rather than quantum mechanical, but is not restricted to rotation – the relations given in Equation (4) hold for any deterministic motion that has a linear generator:

$$\frac{df(\vec{x},t)}{dt} = \frac{\partial f(\vec{x},t)}{\partial t} + \hat{S}f(\vec{x},t) \tag{5}$$

where $f(\vec{x},t)$ is a time dependent function of the coordinates. We do assume, however, that this motion does not depend on spin.

Even with these specifics in place, solving Equation (2) as it stands is a formidable task – it describes unitary probability flow through a space of very high dimension. Our actual target, however, is much simpler – we require the expectation value of the spin density matrix:

$$\hat{\rho}(\vec{x},t) = \int p(\vec{x},\hat{\rho},t)\hat{\rho}\, dV_{\hat{\rho}} \tag{6}$$

where $dV_{\hat{\rho}}$ is the volume element of the density matrix space. The equation of motion for the expectation value may be obtained directly:

$$\frac{\partial}{\partial t}\hat{\rho}(\vec{x},t) = \int\left[\frac{\partial}{\partial t}p(\vec{x},\hat{\rho},t)\right]\hat{\rho}\, dV_{\hat{\rho}} =$$
$$= \int \text{div}_{\hat{\rho}}\left[p(\vec{x},\hat{\rho},t)\left[i\hat{\hat{H}}(\vec{x},t)\hat{\rho}\right]\right]\hat{\rho}\, dV_{\hat{\rho}} + \int \text{div}_{\vec{x}}\left[p(\vec{x},\hat{\rho},t)\left[i\hat{S}\vec{x}\right]\right]\hat{\rho}\, dV_{\hat{\rho}} \tag{7}$$

After a few straightforward vector calculus transformations, we get:

$$\frac{\partial}{\partial t}\hat{\rho}(\vec{x},t) = -i\hat{\hat{H}}(\vec{x},t)\hat{\rho}(\vec{x},t) - i\hat{S}\hat{\rho}(\vec{x},t) \tag{8}$$

In the specific case of spinning at the magic angle:

$$\frac{\partial}{\partial t}\hat{\rho}(\vec{x},t) = -i\hat{\hat{H}}(\vec{x},t)\hat{\rho}(\vec{x},t) - \frac{i\omega_S}{\sqrt{3}}\left[\hat{L}_X + \hat{L}_Y + \hat{L}_Z\right]\hat{\rho}(\vec{x},t) \qquad (9)$$

where $\{\hat{L}_X, \hat{L}_Y, \hat{L}_Z\}$ are angular momentum (not spin) operators acting in lab space:

$$\hat{L}_X = -i\left(y\frac{\partial}{\partial z} - z\frac{\partial}{\partial y}\right); \quad \hat{L}_Y = -i\left(z\frac{\partial}{\partial x} - x\frac{\partial}{\partial z}\right); \quad \hat{L}_Z = -i\left(x\frac{\partial}{\partial y} - y\frac{\partial}{\partial x}\right) \qquad (10)$$

and $\omega_S$ is the spinning frequency. Equation (9) is similar to the Liouville - von Neumann equation: the spin system evolution happens differently at each point in the lab space, but the additional twist is that there is now a coherent flux of spin systems between those points. The primary difference with the SLE formalism[3,6] is that the evolution prescribed by the right hand side of Equation (9) is unitary. Note the minimal modifications that are required to adapt a working SLE code for the Fokker-Planck MAS (FPMAS) formalism: it is a simple matter of taking a different power of the lab space rotation generators. An SLE-type diffusion term may be included if necessary, and so can chemical exchange in a manner identical to the method proposed for Floquet theory[29]. It should also be noted that we did not need the very rigorous derivations through cumulant averages of stochastic processes that were required for the SLE[8,30] – the fact that our lab space evolution is deterministic simplifies matters a great deal because Smoluchowski equation may justifiably be used to obtain Equation (2).

A useful property of Equation (9) that it has in common with SLE and inherits from the more general Equation (2) is that distributions over lab space coordinates are a part of the formalism. This proves to be particularly convenient for the powder averaging operation that is often encountered in solid state NMR spectroscopy – if the lab space basis set is chosen to be Wigner *D*-functions:

$$\hat{H}(\Omega, t) = \sum_{lkm} \mathfrak{D}^{(l)}_{km}(\Omega)\hat{Q}^{(l)}_{km}(t), \qquad \hat{\rho}(\Omega, t) = \sum_{lkm} \mathfrak{D}^{(l)}_{km}(\Omega)\hat{\sigma}^{(l)}_{km}(t) \qquad (11)$$

where $\Omega$ is some parameterization of the rotation group, $\hat{Q}^{(l)}_{km}(t)$ are irreducible components of the Hamiltonian, $\hat{\sigma}^{(l)}_{km}(t)$ are irreducible components of the density matrix and $\mathfrak{D}^{(l)}_{km}(\Omega)$ are Wigner *D*-functions, then the powder average observation operation simply amounts to calculating a scalar product of an observable operator $\hat{O}$ with $\hat{\sigma}^{(0)}_{00}(t)$:

$$\frac{1}{8\pi^2}\int_\Omega \text{Tr}\left[\hat{O}^\dagger \hat{\rho}(t)\right]d\Omega = \frac{1}{8\pi^2}\sum_{lkm}\text{Tr}\left[\hat{O}^\dagger \hat{\sigma}_{km}^{(l)}(t)\right]\int_\Omega \mathfrak{D}_{km}^{(l)}(\Omega)d\Omega = \text{Tr}\left[\hat{O}^\dagger \hat{\sigma}_{00}^{(0)}(t)\right] \quad (12)$$

In other words, the very formidable zoo of spherical grids and integration methods simply disappears and is replaced, as we shall see below, by the altogether more agreeable problem of simply storing a bigger matrix. The only adjustable parameter is the cut-off level $l_{max}$ of the spherical expansion of the density matrix in Equation (11), which is exactly equivalent to the spherical rank level of Lebedev type quadratures[31].

## 3. Numerical implementation

A numerical solution to Equation (9) requires a matrix representation. The spin part of the problem has such a representation by construction. The lab space operators may be converted into matrices by choosing either a discrete grid of points or a basis set of continuous functions. Experiences our colleagues have had with SLE[32,33] indicate that the latter is a better choice – we shall therefore proceed to use the following variable separation ansatz:

$$\hat{H}(\vec{x},t) = \sum_m g_m(\vec{x})\hat{Q}_m(t) \qquad \hat{\rho}(\vec{x},t) = \sum_k g_k(\vec{x})\hat{\sigma}_k(t) \quad (13)$$

where $g_m(\vec{x})$ are orthonormal functions of lab space coordinates, $\hat{Q}_m(t)$ are known spin operators and $\hat{\sigma}_k(t)$ are spin operators that should be solved for. The choice of the basis $\{g_m(\vec{x})\}$ creates a matrix representation for the lab space dynamics operator:

$$S_{nk} = \int g_n^*(\vec{x})\hat{S}g_k(\vec{x})dV_{\vec{x}} \quad \Rightarrow \quad \hat{S}g_k(\vec{x}) = \sum_n S_{nk}g_n(\vec{x}) \quad (14)$$

and Equation (9) becomes:

$$\sum_k g_k \frac{\partial \hat{\sigma}_k}{\partial t} = -i\sum_{nk} g_n g_k \hat{\hat{Q}}_n \hat{\sigma}_k - i\sum_{nk} S_{nk} g_n \hat{\sigma}_k \quad (15)$$

in which the products of lab space functions may be expressed as their linear combinations:

$$g_n g_k = \sum_m c_{nkm} g_m \quad \Rightarrow \quad c_{nkm} = \int g_n(\vec{x})g_k(\vec{x})g_m^*(\vec{x})dV$$

$$\sum_k g_k \frac{\partial \hat{\sigma}_k}{\partial t} = -i\sum_{nkm} c_{nkm} g_m \hat{\hat{Q}}_n \hat{\sigma}_k - i\sum_{nk} S_{nk} g_n \hat{\sigma}_k \quad (16)$$

Taking the scalar product on both sides with each spatial basis function $g_m(\vec{x})$ in turn yields the following system of equations:

$$\left\{\frac{\partial \hat{\sigma}_m}{\partial t}\right\}_m = -i\sum_{nk} c_{nkm} \hat{\hat{Q}}_n \hat{\sigma}_k - i\sum_k S_{mk}\hat{\sigma}_k \tag{17}$$

Finally, after collecting some terms, we get the following system of matrix equations:

$$\left\{\frac{\partial \hat{\sigma}_m}{\partial t}\right\}_m = -i\sum_k \left[\hat{\hat{H}}_{km} + S_{mk}\hat{\hat{E}}\right]\hat{\sigma}_k, \qquad \hat{\hat{H}}_{km} = \sum_n c_{nkm}\hat{\hat{Q}}_n \tag{18}$$

where $\hat{\hat{E}}$ is the identity superoperator in the spin space. These matrix equations may be efficiently solved in the time domain using Krylov propagation[34] or in the frequency domain by calculating the Laplace transform using the Lanczos algorithm[3].

A particularly convenient practical choice for the lab space basis is Wigner *D*-functions. The associated spin operators are known as irreducible spherical tensors[3-6]. Using Equations (11) for the spherical tensor expansion of the Hamiltonian and the density matrix, we obtain:

$$\hat{H}(\Omega,t) = \hat{H}_{\text{ISO}} + \sum_{km} \mathfrak{D}^{(2)}_{km}(\Omega)\hat{Q}_{km}(t), \qquad \hat{\rho}(\Omega,t) = \sum_{lkm}\mathfrak{D}^{(l)}_{km}(\Omega)\hat{\sigma}^{(l)}_{km}(t) \tag{19}$$

where $\hat{H}_{\text{ISO}}$ is the orientation-independent part of the Hamiltonian, $\hat{Q}_{km}(t)$ are irreducible components of its anisotropic part and $\Omega$ is a parameterization of the rotation group, in practice usually Euler angles. Because the anisotropic Hamiltonian only contains Wigner *D*-functions of the second rank, only $l_1 = 2$ structure constants are in practice required:

$$\mathfrak{D}^{(2)}_{km}(\Omega)\mathfrak{D}^{(l)}_{pq}(\Omega) = \sum_{L=|l-2|}^{l+2}\sum_{MN} C^{L,M}_{2,k,l,p} C^{L,N}_{2,m,l,q}\mathfrak{D}^{(L)}_{MN}(\Omega) \tag{20}$$

This constraint makes the evaluation of Clebsch-Gordan coefficients $C^{L,M}_{l_1,m_1,l_2,m_2}$ very affordable because simple analytical expressions exist for the special case of $l_1 = 2$. After matrix representations are obtained, Equation (9) acquires the following general algebraic form:

$$\frac{d}{dt}\vec{\rho} = -i\left[\mathbf{E}\otimes\mathbf{H}_{\text{ISO}} + \sum_{km}\mathbf{D}^{(2)}_{km}\otimes\mathbf{Q}_{km} + \mathbf{S}\otimes\mathbf{E}\right]\vec{\rho} \tag{21}$$

where $\mathbf{E}$ is the unit matrix, $\mathbf{H}_{\text{ISO}}$ is the isotropic Hamiltonian commutation superoperator matrix, $\mathbf{D}^{(2)}_{km}$ are matrix representations of right-sided Wigner function multiplication operators, $\mathbf{Q}_{km}$ are the irreducible components of the anisotropic part of the Hamiltonian commutation superoperator matrix, $\mathbf{S}$ is the spatial dynamics operator matrix and $\vec{\rho}$ is a vector representation of the density operator. In Equation (21) matrices on the right side of Kronecker products are representations of spin operators and those on the left side are representations of operators acting on the lab space degrees of freedom.

In practical calculations, Equation (20) is used to generate matrix representations for the Wigner *D*-function multiplication operators, Equation (19) is used to partition the system into rank and projection blocks and the calculation is carried out forward in time using standard matrix exponentiation techniques in the Kronecker product of lab and spin spaces. SLE and FPMAS modules of our *Spinach* library[9] contain a documented open-source *Matlab* implementation of this entire section as well as source code for the practical simulation examples given below. As is frequently the case with theoretical papers and *Matlab*, the actual program code is shorter than the text of this section.

The dimension of the matrix in Equation (17) can be large (Table 1 and Figure 1) – it is a product of the dimensions of the spin space and the lab space. Unless special measures are taken[22,23], the former grows exponentially with the number of spins in the system and the latter grows cubically as a function of the spherical rank cut-off level $l_{\max}$. For the spin systems and Wigner *D*-function ranks required in practical simulations of common spin systems this dimension is on the brink of capability of modern computers (Table 1). An elegant way around this problem is to compute matrix-vector products prescribed by Equation (21) without evaluating the direct products. This is always possible – given matrices **A** and **B** with elements $A_{n,m}$ and $B_{p,q}$ acting in spin space and lab space respectively, and a vector $\vec{\rho}$ that is represented in the direct product of these spaces, the matrix-vector product $\vec{\sigma} = (\mathbf{A} \otimes \mathbf{B})\vec{\rho}$ can be computed without forming $\mathbf{A} \otimes \mathbf{B}$ explicitly:

$$\sigma_{n,p} = \sum_{m,q} \left[ A_{n,m} B_{p,q} \right] \rho_{m,q} = \sum_{m,q} A_{n,m} \rho_{m,q} B_{p,q} \qquad (22)$$

Here elements of vectors $\vec{\sigma}$ and $\vec{\rho}$ depend on both spin indices $n,m$ and lab space indices $p,q$. In matrix notation this equation is simply **σ = AρB**, where **σ** and **ρ** are the vectors $\vec{\sigma}$ and $\vec{\rho}$ reshaped into spin-by-lab matrices. Such reshaping does not require any permutations of the arrays stored in memory, it only switches the way that these arrays are seen: for instance, **σ** is a matrix with spin and lab indices being row and column indices and $\vec{\sigma}$ is a vector stored along a single multi-index $i = (n-1)P + p$, where $P$ denotes the number of possible values for $p$. With this in place, the matrix-vector multiplication on the right hand side of Equation (17) can be performed as a sum of two-sided matrix multiplications that avoid the calculation of the very large $\mathbf{A} \otimes \mathbf{B}$ matrices. Because Krylov time propagation algorithms only require matrix-vector multiplication, this means that the time propagation problem can likewise be solved at a much reduced memory cost. More generally, the structure of the matrix in Equation (21) is a simple example of a tensor train, an object that generalizes, from the numerical linear algebra point of view, the MPS/DMRG methods proposed initially in quantum physics[35]. Very efficient algorithms exist for manipulating tensor trains without unfolding the direct products[35,36].

The primary advantage of Equation (22) is in the memory requirements – for storage purposes the product of the numbers of non-zeros in $\mathbf{A}$ and $\mathbf{B}$ matrices is effectively replaced by their sum. This is illustrated in the two rightmost columns of Table 1 – the memory requirements for the storage of a tensor train representation of an FPMAS Liouvillian are about the same as those of single-orientation Floquet run and in the case of sucrose are actually smaller. For sucrose (a 12-spin system for $^{13}$C MAS NMR purposes) this proves to be critical – the sparse matrix representation of the FPMAS Liouvillian crashes our best supercomputer, while the tensor train representation fits into the L3 cache of a single CPU. Given the ubiquity of direct product sums in magnetic resonance theory, it is quite clear that tensor train representations of spin operators should be used more widely in simulations.

## 4. Relation to Floquet theory

Floquet theory, as applied to NMR spectroscopy, is a special case of Equation (8). This relationship becomes clear when the fairly arbitrary lab space motion permitted in the right hand side of Equation (8) is constrained to be uniform rotation around a specific axis:

$$\frac{\partial}{\partial t} \hat{\rho}(\varphi, t) = -i\hat{\hat{H}}(\varphi) \hat{\rho}(\varphi, t) + \omega_S \frac{\partial}{\partial \varphi} \hat{\rho}(\varphi, t) \tag{23}$$

where the orientation is now described by an angle $\varphi$, the Hamiltonian superoperator is assumed to only depend on that angle, $\partial/\partial\varphi$ is the rotation generator and $\omega_S$ is the spinning rate. Because the angle $\varphi$ is now periodic, the following variable separation ansatz may be used without loss of generality:

$$\hat{\rho}(\varphi, t) = \sum_n \hat{\rho}_n(t) \exp(in\varphi) \qquad \hat{H}(\varphi) = \sum_k \hat{H}_k \exp(ik\varphi) \tag{24}$$

where the Fourier indices $n$ and $k$ run over all integers. After the substitution is performed, Equation (23) becomes:

$$\sum_{n} \exp(in\varphi)\left[\frac{\partial}{\partial t}\hat{\rho}_n(t)\right] = \sum_{kn} \exp[i(k+n)\varphi]\left[-i\hat{H}_k\hat{\rho}_n(t)\right] + \sum_{n} \exp(in\varphi)\left[in\omega_S\hat{\rho}_n(t)\right] \quad (25)$$

After rotating the summation indices of the double sum and equating the coefficients of the same Fourier factors, we obtain:

$$\left\{\frac{\partial}{\partial t}\hat{\rho}_m(t) = -i\sum_{n}\hat{H}_{m-n}\hat{\rho}_n(t) + im\omega_S\hat{\rho}_m(t)\right\}_m \quad (26)$$

which is identical to the Floquet theory equations for the Fourier components of the density matrix[10-13]. This completes the proof and has a side effect of providing a rather neat alternative derivation for the Floquet MAS formalism.

The primary advantage of Equation (8) over Equation (26) is that three-angle powder averages are built into the FPMAS formalism – as demonstrated in Equation (12), there is no need for a spherical quadrature grid. Another advantage is greater variety of spatial dynamics models that the right hand side of Equation (8) can accommodate – the $\hat{S}$ term may be set to any linear operator, including also translation and diffusion[3]. That having been said, Floquet theory does hold its own in one significant category – the ease of parallelization: it is quite straightforward to distribute several thousand independent Floquet calculations for different system orientations to different nodes of a cluster, but considerably harder (though not impossible[37]) to parallelize a single-trajectory simulation of the kind required by the FPMAS formalism.

## 5. Incomplete basis sets

A general observation in liquid state magnetic resonance spectroscopy is that the complete basis set in the spin operator space is rarely necessary because amplitudes of highly correlated spin states are kept low by the inevitable presence of relaxation processes[23,38]. Very fast simulation algorithms are obtained when unpopulated spin states are dropped from the basis set[25,39]. The same is not in general true for solid state NMR, where dense networks of strong interactions populate high correlation orders in milliseconds – the left panel of Figure 2 demonstrates this for alanine during a proton-decoupled MAS pulse-acquire experiment on

$^{13}$C at a single randomly selected orientation. From the immediate appearance of the graph one might conclude that state space restriction approximation is not applicable to solid state NMR – all spin correlation orders get populated quickly to levels that clearly cannot be ignored.

We would like to report a surprising observation though – as the right panel of Figure 2 demonstrates, effective amplitudes of multi-spin correlations are much reduced when a powder average of the density matrix is taken at each point in the trajectory. In practice that means that *even though highly correlated states are active in each individual crystallite, the spectrometer coil does not feel their presence.* For the purposes of simulating spinning powder-averaged NMR spectra the high spin correlations can therefore be ignored, implying that polynomially scaling algorithms of the same kind as were developed for liquids[9,22,23,39] could be possible, at least for pulse-acquire MAS NMR experiments. A quick look at the simulation results in a simple spin system (Figure 3) confirms this – the simulated spectra with complete and truncated basis sets are essentially the same across the range of spinning rates. Another example, along with experimental data, is given in Figure 4 for uniformly $^{13}$C and $^{15}$N labelled tryptophan – a large spin system not at present accessible to a complete basis set simulation. Even though three-spin and higher correlations are clearly present in the quantum trajectory of each individual crystallite, accurate answers are still obtained when they are ignored. This phenomenon might be general – a similar finding was reported in the recent work on spin diffusion by the Emsley group[25,40,41], who also found that powder averaging mysteriously improved the accuracy of a low correlation order basis set that was not, from common sense expectations, supposed to work at all.

A qualitative explanation as to why reduced basis sets suffice for the simulation of powder MAS spectra may be obtained by inspection of the effect that the powder averaging has on the spin system propagator of a stationary sample. For the free induction decay $f(t)$ at time $t$ of a system that has started evolution in a state $|\hat{\rho}(0)\rangle$ and is detected at the state $|\hat{\sigma}\rangle$, we have:

$$f(t) = \frac{1}{8\pi^2} \int_\Omega \left[ \langle \hat{\sigma} | \exp\left(-i\hat{\hat{H}}(\Omega)t\right) | \hat{\rho}(0) \rangle \right] d\Omega \qquad (27)$$

Using the irreducible spherical tensor expansion for the Hamiltonian:

$$\hat{H}(\Omega) = \sum_{lkm} \hat{Q}_{km}^{(l)} \mathfrak{D}_{km}^{(l)}(\Omega) \qquad (28)$$

and a Taylor series expansion for the matrix exponential, we get:

$$f(t) = \frac{1}{8\pi^2} \int_\Omega \left[ \langle \hat{\sigma} | \sum_{n=0}^{\infty} \frac{(-it)^n}{n!} \left( \sum_{lkm} \hat{\hat{Q}}^{(l)}_{km} \mathfrak{D}^{(l)}_{km}(\Omega) \right)^n | \hat{\rho}(0) \rangle \right] d\Omega = $$
$$= \frac{1}{8\pi^2} \sum_{n=0}^{\infty} \frac{(-it)^n}{n!} \left[ \langle \hat{\sigma} | \int_\Omega \left( \sum_{lkm} \hat{\hat{Q}}^{(l)}_{km} \mathfrak{D}^{(l)}_{km}(\Omega) \right)^n d\Omega | \hat{\rho}(0) \rangle \right]$$
(29)

The initial and the detection state in MAS NMR contain only single-spin correlations and the question is therefore narrowed to which correlation orders are populated and subsequently returned into the single-spin order subspace by the following superoperators:

$$\frac{1}{8\pi^2} \int_\Omega \left( \sum_{lkm} \hat{\hat{Q}}^{(l)}_{km} \mathfrak{D}^{(l)}_{km}(\Omega) \right)^n d\Omega = \frac{1}{8\pi^2} \int_\Omega \left( \sum_{lkm} \hat{\hat{P}}^{(l,n)}_{km} \mathfrak{D}^{(l)}_{km}(\Omega) \right) d\Omega$$
(30)

The right hand side of this equation follows from the fact that Wigner *D*-functions are a complete basis set in the space of all continuous functions of $\Omega$. Due to the restrictions on the physical nature of spin interaction operators only $l \leq 2$ ranks appear in $\hat{\hat{Q}}^{(l)}_{km}$, but $\hat{\hat{P}}^{(l,n)}_{km}$ operators can in general have any $l$ rank.

The presence of the full spherical average is now revealed to be the critical factor – most Wigner *D*-functions integrate to zero and only the terms multiplied by $\mathfrak{D}^{(0)}_{00}(\Omega)$ survive the powder integration:

$$\frac{1}{8\pi^2} \int_\Omega \left( \sum_{lkm} \hat{\hat{P}}^{(l,n)}_{km} \mathfrak{D}^{(l)}_{km}(\Omega) \right) d\Omega = \hat{\hat{P}}^{(0,n)}_{00}$$
(31)

The remaining operators $\hat{\hat{P}}^{(0,n)}_{00}$ are complicated linear combinations of powers of $\hat{\hat{Q}}^{(l)}_{km}$ – we could not find a way to express them compactly, but what is clear from Equation (31) is that a massively smaller number of spin operators is present in the effective powder average propagator compared to the single crystal propagator – just one per Taylor series term, compared to the full three-index sum from Equation (30) for the single crystal case. The zero spherical rank and projection quantum numbers of $\hat{\hat{P}}^{(0,n)}_{00}$ also explain why the dynamics mostly stays in the low correlation order subspace. These are qualitative arguments, however – a more rigorous and quantitative analysis has thus far defeated our attempts. It should also be noted that the observations reported in this section apply in equal measure to both Floquet and Fokker-Planck description of MAS NMR – the simulation results are identical.

## 6. Numerical performance

The price for the generality of the motion model and for the removal of spherical grids in the Fokker-Planck formalism is paid in matrix dimensions. Those are astronomical, even by the liberal standards of Floquet theory (Table 1), but the ability to work with large matrices has much improved in recent years – it is currently possible to handle spin Liouvillians with dimensions well in excess of $10^7$, particularly if Lanczos type algorithms[3], sparse libraries[42,43] and state space restriction methods[9,22,23] are used. One saving grace is that the matrices involved are always very sparse – this has been known for a long time[29,43], but we recently also demonstrated that the fraction of non-zeros in spin Hamiltonians drops exponentially as the system size is increased[37]. Another saving grace is the above noted tensor train format for the Liouvillian storage[36]; it does not reduce the amount of multiplications needed, but shrinks the memory footprint of the simulation by several orders of magnitude. It becomes essential for the simulation of systems with more than about five spins.

Because FPMAS solves for the three-angle orientational average, a set of three-angle GSQ grids (which descend from Lebedev grids and whose spherical rank is equivalent to FPMAS Wigner rank[17]) are used with Floquet theory to provide the apples-to-apples performance comparison given in Table 2. The conclusion is that there is no significant consistent performance difference between the two theories – an order of magnitude or so can be seen in favour of either Floquet or Fokker-Planck, the winner depending on the system. When state space restriction techniques are used (*e.g.* only the +1 coherence level is necessary for pulse-acquire simulations, only longitudinal states are populated on $^{15}$N, *etc.*), FPMAS formalism is marginally faster.

## 7. Conclusions

We propose a Fokker-Planck type formalism for the simulation of magic angle spinning NMR experiments. The two primary advantages over the *status quo* are that sample spinning is treated as just another time-independent interaction term in the Liouvillian and powder averages are obtained in a single run – there is no spherical quadrature grid. Relaxation theory with any linear diffusion operator may be easily adopted from the closely related Stochastic Liouville Equation formalism by adding a diffusion term to the right hand side of Equation (8). We have also demonstrated that the Fokker-Planck MAS formalism contains the NMR version of Floquet theory as a special case.

The elimination of the spherical grid comes at the cost of increased matrix dimensions, but we show that this can be mitigated by the use of state space restriction and tensor train techniques. We have also demonstrated that low correlation order basis sets apparently give accurate answers in powder MAS simulations, meaning that polynomially scaling simulation algorithms do exist for a large class of solid state NMR experiments.

## Acknowledgements

We are grateful to Jean-Nicolas Dumez, Jack Freed, Meghan Halse, Eva Meirovitch, Malcolm Levitt and Shimon Vega for very useful discussions. The project is supported by EPSRC (EP/H003789/1).

# Literature

(1) Fokker, A. D. *Annalen der Physik* **1914**, *348*, 810.

(2) Planck, M. *Sitzungsber. Kön. Preuss. Akad. Wiss.* **1917**, 324.

(3) Moro, G.; Freed, J. H. *J. Chem. Phys.* **1981**, *74*, 3757.

(4) Mason, R. P.; Freed, J. H. *J. Phys. Chem.* **1974**, *78*, 1321.

(5) Polnaszek, C. F.; Bruno, G. V.; Freed, J. H. *J. Chem. Phys.* **1973**, *58*, 3185.

(6) Freed, J. H.; Bruno, G. V.; Polnaszek, C. F. *J. Phys. Chem.* **1971**, *75*, 3385.

(7) Nevzorov, A. A. *J. Phys. Chem. B* **2011**, *115*, 15406.

(8) Nevzorov, A. A.; Freed, J. H. *J. Chem. Phys.* **2000**, *112*, 1413.

(9) Hogben, H. J.; Krzystyniak, M.; Charnock, G. T. P.; Hore, P. J.; Kuprov, I. *J. Magn. Reson.* **2011**, *208*, 179.

(10) Scholz, I.; van Beek, J. D.; Ernst, M. *Solid State NMR* **2010**, *37*, 39.

(11) Leskes, M.; Madhu, P. K.; Vega, S. *Progr. NMR Spec.* **2010**, *57*, 345.

(12) Bain, A. D.; Dumont, R. S. *Conc. Magn. Reson.* **2001**, *13*, 159.

(13) Schmidt, A.; Vega, S. *The Journal of Chemical Physics* **1992**, *96*, 2655.

(14) Stevensson, B.; Edén, M. *J. Magn. Reson.* **2006**, *181*, 162.

(15) Edén, M. *Conc. Magn. Reson.* **2003**, *18*, 24.

(16) Ponti, A. *J. Magn. Reson.* **1999**, *138*, 288.

(17) Edén, M.; Levitt, M. H. *J. Magn. Reson.* **1998**, *132*, 220.

(18) Bak, M.; Nielsen, N. C. *J. Magn. Reson.* **1997**, *125*, 132.

(19) Varner, S. J.; Vold, R. L.; Hoatson, G. L. *J. Magn. Reson.* **1996**, *123*, 72.

(20) Wang, D.; Hanson, G. R. *J. Magn. Reson.* **1995**, *117*, 1.

(21) Mombourquette, M. J.; Weil, J. A. *J. Magn. Reson.* **1992**, *99*, 37.

(22) Kuprov, I. *J. Magn. Reson.* **2008**, *195*, 45.

(23) Kuprov, I.; Wagner-Rundell, N.; Hore, P. J. *J. Magn. Reson.* **2007**, *189*, 241.

(24) Karabanov, A.; van der Drift, A.; Edwards, L. J.; Kuprov, I.; Kockenberger, W. *Phys. Chem. Chem. Phys.* **2012**, *14*, 2658.

(25) Butler, M. C.; Dumez, J. N.; Emsley, L. *Chem. Phys. Lett.* **2009**, *477*, 377.

(26) Levitt, M. H. *Spin dynamics: basics of nuclear magnetic resonance*; Wiley, 2008.

(27) Ernst, R. R.; Bodenhausen, G.; Wokaun, A. *Principles of nuclear magnetic resonance in one and two dimensions*; Oxford University Press, 1987.

(28) Abragam, A. *The principles of nuclear magnetism*; Clarendon, 1961.

(29) Hazendonk, P.; Bain, A. D.; Grondey, H.; Harrison, P. H. M.; Dumont, R. S. *J. Magn. Reson.* **2000**, *146*, 33.

(30) Pedersen, J. B. In *Electron Spin Relaxation in Liquids*; Plenum: 1972, p 25.

(31) Lebedev, V. I.; Laikov, D. N. *Doklady Akademii Nauk* **1999**, *366*, 741.

(32) Stillman, A. E.; Zientara, G. P.; Freed, J. H. *J. Chem. Phys.* **1979**, *71*, 113.

(33) Freed, J. H.; Pedersen, J. B. *Adv. Magn. Reson.* **1976**, *8*, 2.

(34) Tannor, D. J. *Introduction to quantum mechanics: a time-dependent perspective*; University Science Books, 2007.

(35) Schollwock, U. *Ann. Phys.* **2011**, *326*, 96.

(36) Oseledets, I. V. *SIAM J. Sci. Comp.* **2011**, *33*, 2295.

(37) Edwards, L. J.; Kuprov, I. *J. Chem. Phys.* **2012**, *136*.

(38) Karabanov, A.; Kuprov, I.; Charnock, G. T. P.; van der Drift, A.; Edwards, L. J.; Koeckenberger, W. *J. Chem. Phys.* **2011**, *135*.

(39) Hogben, H. J.; Hore, P. J.; Kuprov, I. *J. Chem. Phys.* **2010**, *132*.

(40) Dumez, J. N.; Halse, M. E.; Butlerz, M. C.; Emsley, L. *Phys. Chem. Chem. Phys.* **2012**, *14*, 86.

(41) Dumez, J. N.; Butler, M. C.; Emsley, L. *J. Chem. Phys.* **2010**, *133*.

(42) Veshtort, M.; Griffin, R. G. *J. Magn. Reson.* **2006**, *178*, 248.

(43) Dumont, R. S.; Jain, S.; Bain, A. *J. Chem. Phys.* **1997**, *106*, 5928.

(44) Becke, A. D. *J. Chem. Phys.* **1993**, *98*, 1372.

(45) Dunning, T. H. *J. Chem. Phys.* **1989**, *90*, 1007.

(46) Lee, C. T.; Yang, W. T.; Parr, R. G. *Phys. Rev. B* **1988**, *37*, 785.

(47) London, F. *J. Phys. Radium* **1937**, *8*, 397.

# Figure captions

**Figure 1.** Results and matrix dimension statistics for basic CSA, dipolar and quadrupolar powder patterns calculated for different sample spinning rates with Floquet and Fokker-Planck formalism. The following algorithm parameters were used, selected in such a way as to achieve numerical convergence of the resulting spectra to approximately $10^{-3}$ relative accuracy: *(10 kHz trace)* maximum Fourier rank 5, GSQ grid rank 7 for Floquet simulation, maximum Wigner function rank 7 for Fokker-Planck simulation; *(1.0 kHz trace)* maximum Fourier rank 10, GSQ grid rank 17 for Floquet simulation, maximum Wigner function rank 17 for Fokker-Planck simulation; *(0.1 kHz trace)* maximum Fourier rank 40, GSQ grid rank 35 for Floquet simulation, maximum Wigner function rank 35 for Fokker-Planck simulation. Three-angle GSQ grids[17] on the Floquet side and full Wigner function basis sets on the Fokker-Planck side were chosen to provide an apples-to-apples comparison: two-angle averages would have sufficed for the simulations shown above, but we have chosen to use the most general simulation case for the comparison.

**Figure 2.** Spin system trajectory analysis during the detection period of $^{13}$C pulse-acquire MAS NMR experiment on $^{13}$C,$^{15}$N-labelled alanine under proton decoupling. Left panel – spin correlation order dynamics in a single crystallite: the state space is filled completely with significant population in four-spin orders. Right panel – spin correlation order dynamics in a powder-averaged density matrix: orientation averaging appears to reduce the effective level of spin correlation in the trajectory, the populations of three- and four-spin orders are significantly reduced.

**Figure 3.** Basis set convergence of restricted state space calculations of $^{13}$C pulse-acquire MAS NMR spectra of $^{13}$C,$^{15}$N-labeled alanine powder under proton decoupling. The difference between the full state space (top row) and reduced state spaces (middle and bottom rows) is minor – the maximum relative difference from the exact simulation is ~$10^{-3}$ for three-spin order basis set and ~$10^{-2}$ for the two-spin order basis set. This level of accuracy is unexpected because three- and four-spin correlations do get populated (see Figure 2) during the system evolution.

**Figure 4.** Basis set convergence and comparison with the experimental data for the restricted state space calculations of $^{13}$C pulse-acquire MAS NMR spectra of uniformly $^{13}$C,$^{15}$N-labeled tryptophan powder under proton decoupling. For the purposes of the simulation, Cartesian coordinates of the tryptophan amino acid were extracted from the Cambridge Structural Database, the coordinates of the hydrogen atoms were re-optimized using DFT B3LYP/cc-pVTZ method[44-46] *in vacuo*. Isotropic chemical shifts were read off from the experimental data and chemical shielding tensor anisotropies were estimated with GIAO B3LYP/cc-pVTZ method[47] and used without further fitting. *J*-couplings were estimated using the same method with a decontracted cc-pVTZ basis set augmented with tight functions in the nuclear region. All DFT calculations were performed in Gaussian09. Experimental $^1$H-decoupled CP-MAS spectra were recorded for a powdered sample of uniformly $^{13}$C,$^{15}$N-labeled D,L-tryptophan (Cambridge Isotope Labs) packed in a 4 mm zirconia rotor. A 400 MHz Varian Infinity Plus NMR spectrometer was used. Spectra were acquired with three different values of the spinning rate: 14, 10, 7 kHz (top to bottom). SPINAL decoupling was used with a proton pulse duration of 6.5 µs corresponding to a nutation frequency of 75 kHz. Cross-polarization conditions were adjusted for each spinning rate value; for the 10 kHz trace a contact time of 2.5 ms and a nutation frequency of 50 kHz for protons and 40 kHz for carbon were used.

FIGURE 1

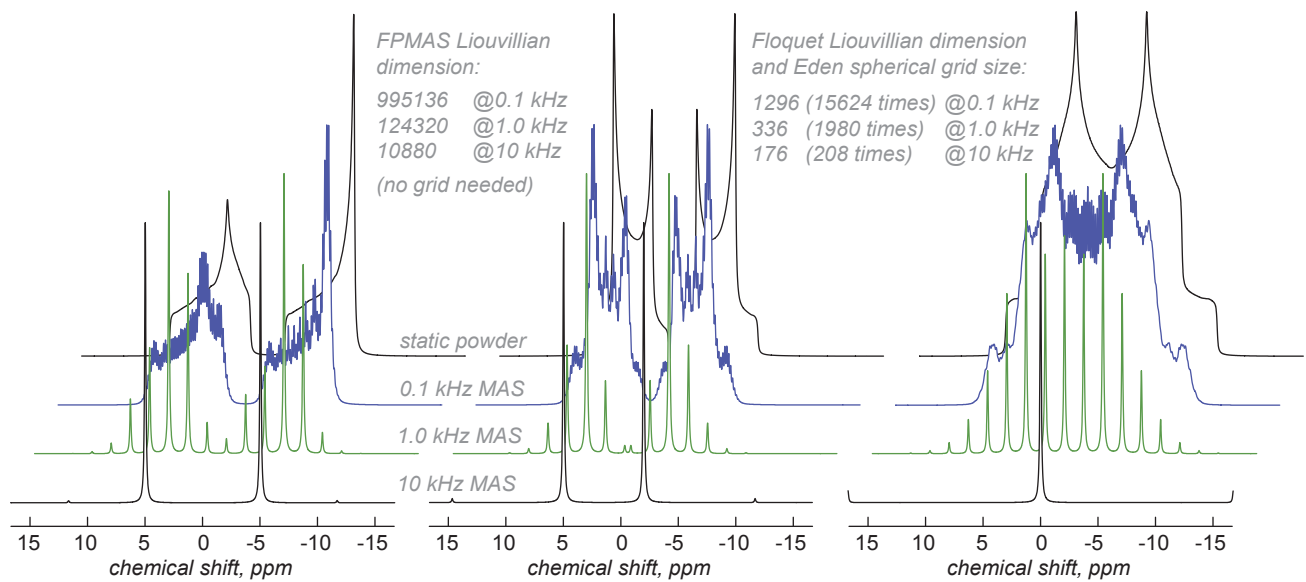

FIGURE 2

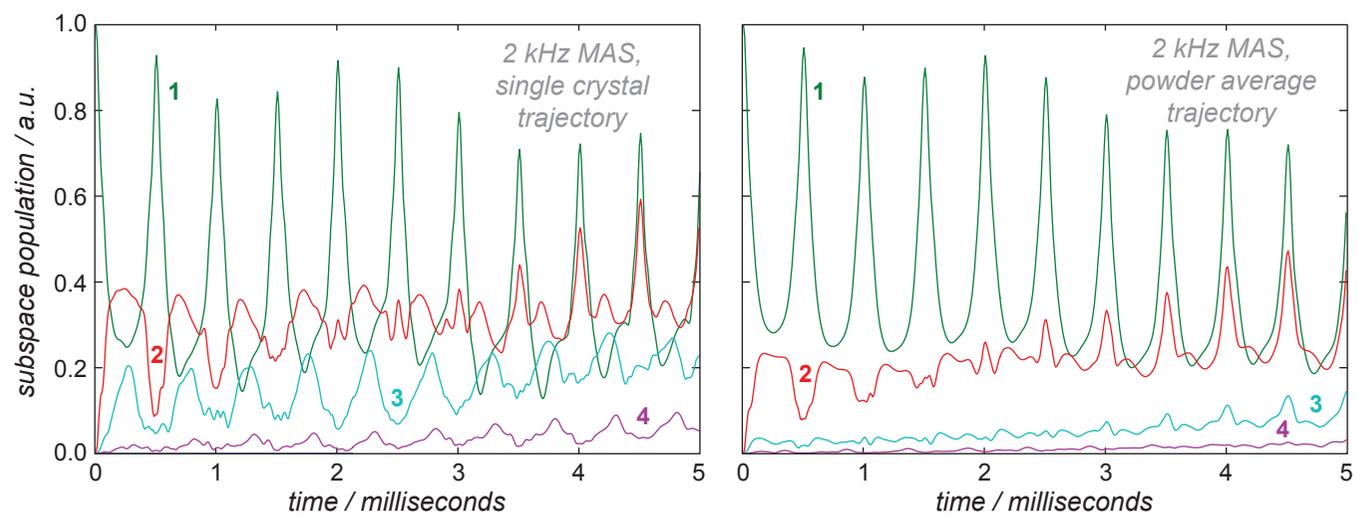

FIGURE 3

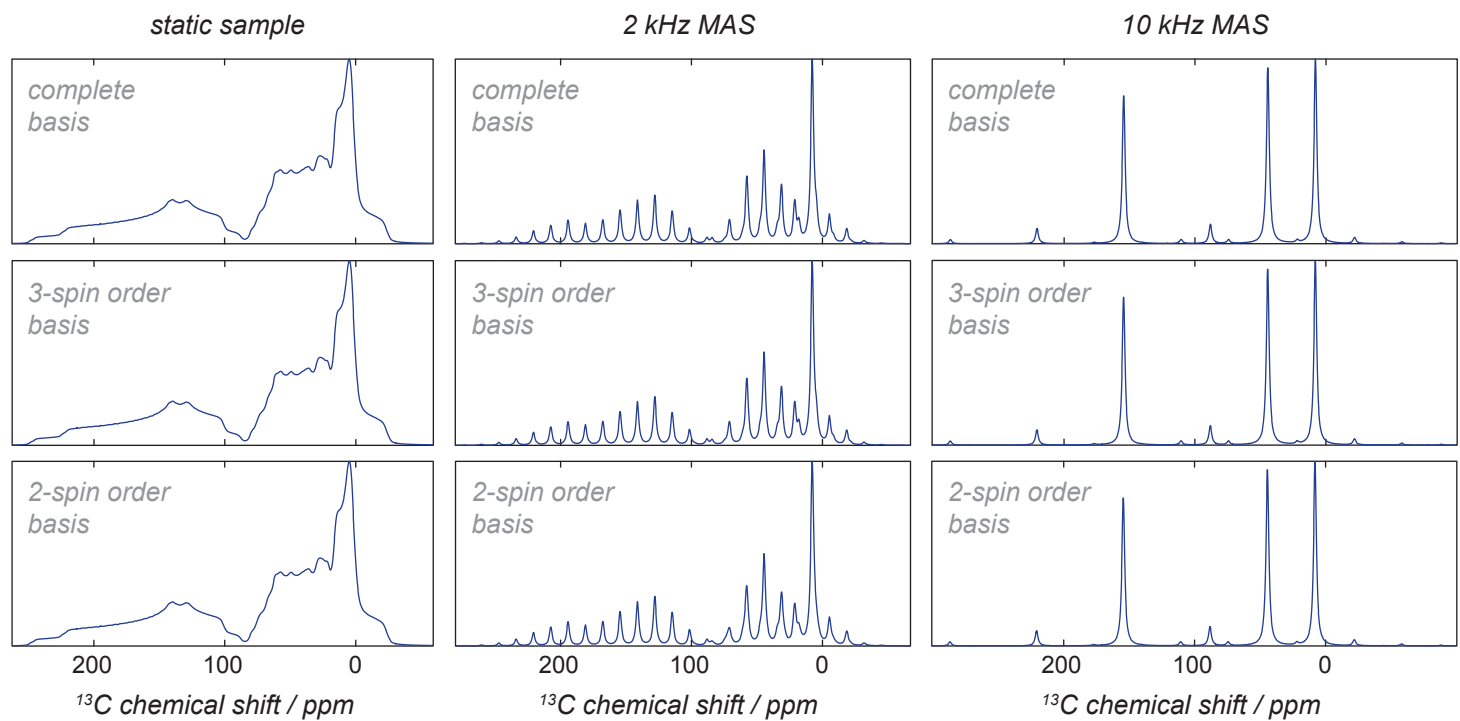

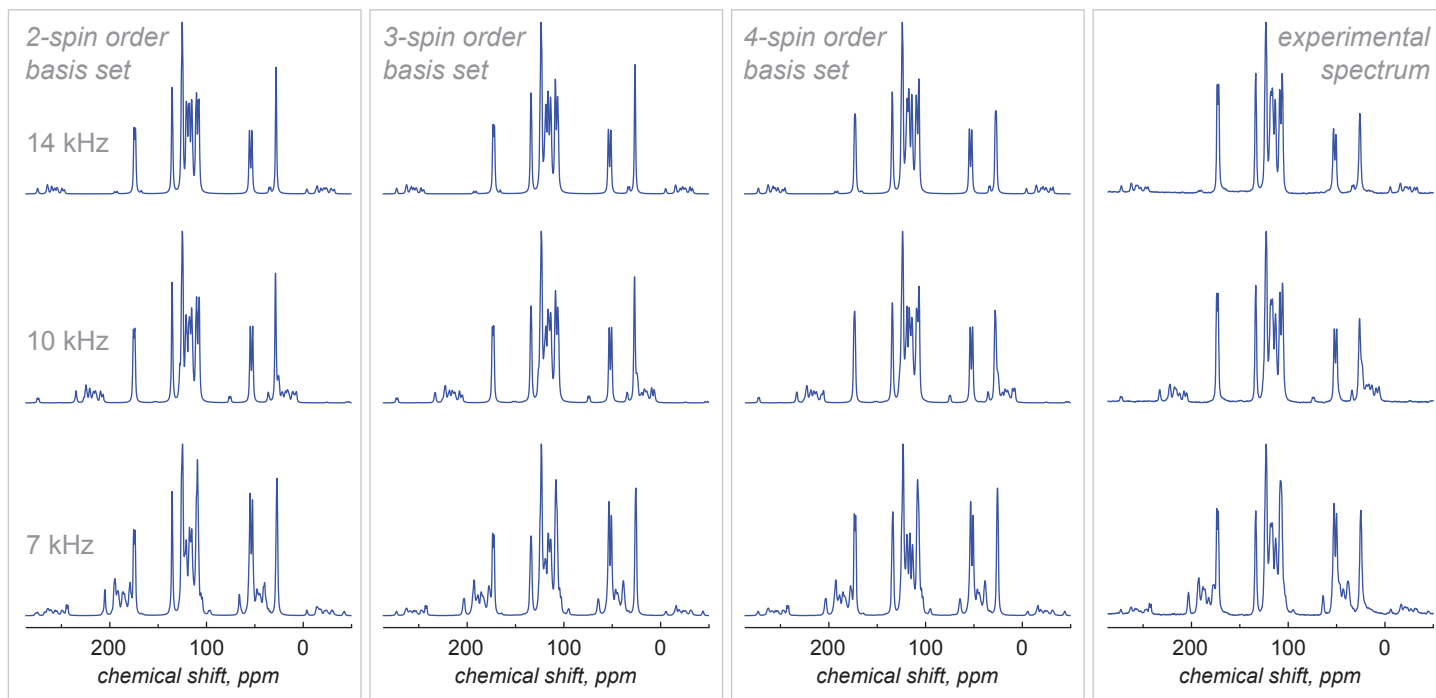

**Table 1**: Liouvillian dimension and non-zero count statistics for glycine, alanine and sucrose spin systems in a proton-decoupled $^{13}$C pulse-acquire 2.0 kHz MAS NMR simulation using Floquet, FPMAS and tensor train FPMAS formalisms.

| System | | Single-point | Floquet on 2-angle grid [a] | Floquet on 3-angle grid [b] | FPMAS, full | FPMAS, reduced [c] | TT-FPMAS [d], full | TT-FPMAS [d], reduced [e] |
|---|---|---|---|---|---|---|---|---|
| Glycine ($^{13}$C,$^{15}$N), Wigner rank 23, Fourier rank 20, complete basis. | *dim* | 64 | 2624 (194 points) | 2624 (4,656 points) | 1,179,136 | 165,816 | 64 ⊗ 18,424 (26 terms) | 9 ⊗ 18,424 (7 terms) |
| | *nnz* | 0.2k | 38k (194 points) | 38k (4,656 points) | 76M | 13M | 456k | 455k |
| Alanine, ($^{13}$C,$^{15}$N), Wigner rank 23, Fourier rank 20, complete basis. | *dim* | 256 | 10,496 (194 points) | 10,496 (4,656 points) | 4,716,544 | 571,144 | 256 ⊗ 18,424 (26 terms) | 31 ⊗ 18,424 (7 terms) |
| | *nnz* | 1.4k | 290k (194 points) | 290k (4,656 points) | 567M | 79M | 464k | 456k |
| Sucrose ($^{13}$C), Wigner rank 23, Fourier rank 20, 3-spin order basis. | *dim* | 6,571 | 269,411 (194 points) | 269,411 (4,656 times) | 121,964,104 | 26,991,160 | 6,571 ⊗ 18,424 (26 terms) | 1,465 ⊗ 18,424 (7 terms) |
| | *nnz* | 49k | 7.1M (194 points) | 7.1M (4,656 points) | >45G | 3.1G | 672k | 507k |

[a] Rank 23 Lebedev quadrature [31]; [b] Rank 23 GSQ quadrature [17]; [c] Conservation law screening [39] followed by zero track elimination [22] followed by path tracing [39]; [d] Tensor train formalism [36]; [e] Conservation law screening [39].

**Table 2**: Floquet and Fokker-Planck simulation run time statistics for glycine, alanine and sucrose spin systems in a proton-decoupled $^{13}$C pulse-acquire 2.0 kHz MAS NMR simulation.

| System | Run time, seconds | | | | | | |
|---|---|---|---|---|---|---|---|
| | Single crystal | Floquet on 2-angle grid[a] | Floquet on 3-angle grid[b] | FPMAS, full | FPMAS, reduced[c] | TT-FPMAS[d], full | TT-FPMAS[d], reduced[e] |
| Glycine ($^{13}$C,$^{15}$N), Wigner rank 23, Fourier rank 20, complete basis. | $2.60 \cdot 10^{-1}$ | $5.04 \cdot 10^{1}$ | $1.21 \cdot 10^{3}$ | $6.25 \cdot 10^{3}$ | $9.21 \cdot 10^{2}$ | $3.58 \cdot 10^{4}$ | $4.20 \cdot 10^{3}$ |
| Alanine, ($^{13}$C,$^{15}$N), Wigner rank 23, Fourier rank 20, complete basis. | $2.89 \cdot 10^{0}$ | $5.60 \cdot 10^{2}$ | $1.34 \cdot 10^{4}$ | $6.14 \cdot 10^{4}$ | $5.69 \cdot 10^{3}$ | $1.84 \cdot 10^{5}$ | $1.82 \cdot 10^{4}$ |
| Sucrose ($^{13}$C), Wigner rank 23, Fourier rank 20, 3-spin order basis. | $5.27 \cdot 10^{2}$ | $1.02 \cdot 10^{5}$ | $2.45 \cdot 10^{6}$ | $>10^{7}$ | $6.24 \cdot 10^{5}$ | $3.39 \cdot 10^{6}$ | $5.83 \cdot 10^{5}$ |

[a] Rank 23 Lebedev quadrature [31]; [b] Rank 23 GSQ quadrature [17]; [c] Conservation law screening [39] followed by zero track elimination [22] followed by path tracing [39]; [d] Tensor train formalism [36]; [e] Conservation law screening [39].